\documentclass[journal]{IEEEtran}
\usepackage{amsmath}
\usepackage{amsfonts}
\usepackage{amssymb}
\usepackage[]{graphicx}
\usepackage{bm}
\usepackage{subfigure}
\usepackage{ragged2e}
\usepackage{setspace}
\usepackage{color}
\usepackage{mathrsfs} 
\usepackage[noadjust]{cite}

\ifCLASSINFOpdf
\else
\fi

\begin{document}
%
\title{Wideband Source Enumeration Using Sparse Array Periodogram Averaging in Low Snapshot Scenarios}
%
%
%


\author{Yang~Liu,~
        John~R~Buck,~\IEEEmembership{Senior Member,~IEEE}
\thanks{Dr. Yang Liu is with the Consumer Electronics Division, Bose Corporation, 100 the Mountain Rd, Framingham, MA  01701 USA (e-mail: yangliu$\_$acoustics@outlook.com). Dr. John R . Buck is with the Department of Electrical and Computer Engineering, University of Massachusetts Dartmouth, 285 Old Westport Rd, Dartmouth,  MA 02747 USA (e-mail: jbuck@umassd.edu). This material is based upon research supported by the U.S. Office of Naval Research under award numbers N00014-13-1-0230, N00014-17-1-2397 and N00014-18-1-2415.}}
\maketitle

\begin{abstract}
This paper proposes a new sparse array source enumeration algorithm for underdetermined scenarios with more sources than sensors. The proposed algorithm decomposes the wideband signals into multiple uncorrelated frequency bands, computes the narrowband spatial periodograms and then averages periodograms to reinforce the sources' spectral information. The inverse spatial Fourier transform of the wideband periodogram estimates the spatial correlation function, which then populates the diagonals of a Hermitian Toeplitz augmented covariance matrix (ACM) after lag redundancy averaging. A modified minimum description length (MDL) criteria, termed MDLgap, is proposed for source enumeration using the eigenvalues of the constructed ACM. MDLgap provably never overestimates the number of sources present, and is asymptotically consistent when the signals present span a limited dynamic range. Numerical simulations show that the proposed MDLgap algorithm achieves improved performance over existing approaches for underdetermined source enumeration, especially in low snapshot scenarios. 
\end{abstract}


%
\IEEEpeerreviewmaketitle

\section{Introduction}
%
%
%
%


\IEEEPARstart{T} \normalsize hinning sensors from a uniform line array (ULA) produces several common families of sparse sensors arrays.  The underlying ULA sensor are typically spaced by half a wavelength for the design frequency of interest
\cite{Johnson}. Sparse arrays such as minimum redundancy arrays (MRA), coprime arrays and nested arrays have many array processing applications including source detection, Direction-of-Arrival (DOA) estimation and spatial power spectral estimation \cite{MRA, PPCSA, PPNested,CSADetectionICASSP,kaushallyaECSA,YangSAM2016}. Assuming a large number of snapshots, sparse array processing techniques can localize more sources than sensors by exploiting the second-order statistics of the propagating electromagnetic or acoustic field \cite{PPNested,PillaiACM,PPCSAMUSIC,Aminmultiple}. In this paper, we follow the practice of calling such scenarios \textit{underdetermined}, because the number of sources exceeds the number of sensors. In many array processing applications such as DOA estimation, the number of sources is often assumed known \textit{a priori} \cite{PPNested,PPCSAMUSIC}, although in practice it must first be estimated from the data. In this regard, the accuracy of the source enumeration algorithm is vital to the subsequent sparse array DOA estimation algorithms. Many source enumeration algorithms have been proposed for data uniformly sampled in time and space, pioneered by the Akaike information criterion (AIC \cite{AIC}) and Rissanen's minimum description length criterion (MDL \cite{MDL}), which both posed the enumeration problem as a model selection problem. The AIC and MDL criteria estimate the number of sources using the eigenvalues of the Wishart-distributed sample covariance matrix (SCM) from a ULA \cite{WaxKailathSN}. However, the augmented covariance matrices (ACM) in sparse array DOA estimation algorithms \cite{PillaiACM,PPCSA, PPNested} are not Wishart distributed. This paper considers the problem of source enumeration for sparse arrays, specifically in the context of exploiting ACMs for underdetermined DOA estimation in snapshot limited scenarios. 


Underdetermined source enumeration using sparse sensor arrays is a challenging and open research topic for several reasons. First, it requires constructing an ACM from the estimated signal spatial correlations to fully exploit the degrees-of-freedom (DOFs) offered by the co-array of the sparse sensor array \cite{PillaiACM,PPCSAMUSIC,PPLiuSPL}. Constructing the ACM through either spatial smoothing \cite{PPNested} or directly populating its diagonals \cite{PPLiuSPL} destroys the Wishart characteristic of the covariance matrix. As a result, the classical source enumeration criteria are no longer applicable. Second, nearly all proposed source enumeration and DOA estimation algorithms on sparse arrays assume large numbers of snapshots to ensure more accurate estimates of the spatial correlation and also build up the rank of the ACM \cite{PPCSA,PPNested,PPCSAMUSIC,Aminmultiple,PPLiuSPL}. While this large number of snapshots are commonly available in electromagnetic applications, they are unrealistically large for many acoustic array processing scenarios due to the slower field propagation speed, large array aperture and the non-stationary field \cite{BragCox,Cox}. For source enumeration using relatively few snapshots, the information criteria developed from random matrix theory (RMT) achieved improved performance over the original AIC and MDL criteria \cite{RajAIC, kritchman2008}. However, the RMT criteria were derived for the eigenvalues of a large-dimensional Wishart distributed SCM while explicitly assuming many fewer sources than sensors, or, overdetermined scenarios. Therefore, the RMT criteria do not apply to the underdetermined scenarios and non-Wishart distributed ACMs in this paper  \cite{Wishart}. A more recent algorithm for sparse arrays \cite{HanWidebandSPL} exploits wideband feature of most communication sources and combines detection information criteria across temporal frequency bands for sparse array source enumeration. However, the algorithm still requires a large number of snapshots to achieve accurate estimates. These prior algorithms focus on one or two of the three source enumeration challenges discussed in here. We are not aware of any prior algorithms designed to enumerate sources in underdetermined scenarios in snapshot limited environments.


We propose a new underdetermined source enumeration algorithm for sparse sensor arrays. The source signals are assumed to be wideband in temporal frequency and occupying $M$ frequency bins, denoted by $f_1,...,f_M$. The proposed algorithm exploits the property that the true source peak locations in the narrowband spatial periodograms remain fixed in angle across temporal frequencies, while the grating lobes and sidelobes change their locations. Therefore, as more frequency bands become available, averaging the narrowband periodograms linearly reinforces the sources' spectral information in the wideband periodogram \cite{Hinich}. Constructing the ACM using the correlation estimated from the wideband periodogram improves the enumeration performance, especially in low snapshot scenarios. 

\section{Problem Formulation}

\label{SecSignalModel}
\subsection{Wideband Signal Model}
Assume a sparse linear array with $N$ sensors and $D$ wideband planewave signals impinging on the array from the far field with different angles $\theta_1, \theta_2,...,\theta_D \in [0^o, 180^o]$. The signal received by the $n$th sensor at time $t$ can be modeled as 
\begin{equation}
x_n(t) = \sum_{i = 1}^D s_i(t-\tau_n(\theta_i)) + \text{n}_n(t),~n=1,...,N,
\end{equation} where $\tau_n(\theta_i)$ is the propagation time delay for the $i$th signal arriving at the $n$th sensor and $\text{n}_n(t)$ is the measurement noise at the $n$th sensor. We assume both the signals and noise measured by the sensors are samples of wide-sense stationary and ergodic complex Gaussian processes. The time series at each sensor is divided into $L$ segments, commonly known as snapshots. Applying the discrete Fourier transform (DFT) to each segment forms multiple non-overlapping narrow frequency bands, from which we extract the frequency domain phasors at the frequencies of interest $f_1,...,f_M$. The snapshot duration is assumed much longer than the signal correlation time, such that the different DFT bins are statistically uncorrelated. The vector of DFT coefficients (or phasors) for all $N$ sensors at frequency $f_m$ and $l^{th}$ snapshot has the form 
\begin{equation}
\label{snapshots}
\textbf{x}_l(f_m) = \textbf{A}(f_m)\textbf{s}_l(f_m) + \textbf{n}_l(f_m),
\end{equation}where $\textbf{A}(f_m)$ is the $N\times D$ array manifold matrix at temporal frequency $f_m, m = 1,...,M$. The array manifold corresponding to the $n^{th}$ element and the $i^{th}$ source at frequency $f_m$ is
\begin{equation}
[\textbf{A}(f_m)]_{n,i} = e^{-j(2\pi f_m d_n/c)u_i},
\end{equation}where $d_n$ is the location of the $n$th element with respect to the array phase center, $u_i = \cos(\theta_i)$ is the directional cosine of the $i$th source defined within the visible region $u_i \in [-1,1]$ with $u_i = 0$ indicating broadside, and $c$ is the field propagation speed. The source signal amplitudes are assumed uncorrelated zero-mean and circular complex Gaussians $s_{l,i}(f_m) \sim CN(0,\sigma^2_{i,m})$ and uncorrelated with the noise. The additive noise is assumed zero-mean, white and circular complex Gaussian $\textbf{n} \sim CN(\textbf{0},\sigma_n^2 \textbf{I}_N)$. 

\subsection{Incoherent Wideband Source Enumeration Approach}
Many eigenvalue-based wideband source enumeration algorithms compute the information criteria such as MDL or AIC for each frequency bin, and then average the information over all frequencies \cite{WaxKailathSN, HanWidebandSPL,RajAIC}. These algorithms are usually referred to as the incoherent subspace (ISS) approach. For underdetermined source enumeration with a sparse array, the ISS approach constructs an ACM for each frequency bin $f_m$. We briefly review the data processing procedures in the context of finite snapshots. For any particular frequency bin $f_m$, compute the SCM from the complex phasors data from $L$ snapshots
\begin{equation}
\textbf{R}_{xx,m} = \frac{1}{L}\sum_{l=1}^L \textbf{x}_l(f_m) \textbf{x}_l^H(f_m),
\end{equation}where $(\cdot)^H$ denotes Hermitian transpose. Selecting appropriate entries from the SCM to populate the $(2P-1)\times 1$ spatial correlation vector corresponding to the contiguous region of the difference co-array yields
\begin{equation}
\label{corr}
\textbf{r}_{m}(k) = \frac{1}{\eta (k)} \sum_{(n_1,n_2)~\in~\zeta(k)} \left[ \textbf{R}_{xx,m} \right]_{n_1,n_2}, 
\end{equation} where $[\textbf{R}]_{n_1,n_2}$ takes the $(n_1,n_2)$th element of matrix $\textbf{R}$. The set $\zeta(k)$ collects every pair ($n_1,n_2$) contributing to the difference co-array index $k = n_1-n_2 \in  [1-P, P-1]$ and $\eta(k)$ is the co-array weight function equal to the cardinality of the set $\zeta(k)$. To exploit fully the DOFs offered by the co-array, apply spatial smoothing (SS) to construct a full-rank and positive semi-definite ACM by \cite{PPNested, PPCSAMUSIC}
\begin{equation}
\label{SS-ACM}
\textbf{R}_{ss,m} = \frac{1}{P}\sum_{i=1}^P \textbf{v}_m^i (\textbf{v}_m^i)^H,
\end{equation}where $\textbf{v}_m^i$ is a $P\times 1$ vector containing the ($P-i+1$)th through ($2P-i$)th elements of $\textbf{r}_{m}(k)$. The final step computes the information criteria from the eigenvalues of the SS-ACMs $\textbf{R}_{ss,m}$ for each frequency, and then averages these criteria across frequency bins to estimate the number of sources $D$ \cite{HanWidebandSPL}. While the incoherent source enumeration approach works relatively well for wideband sources in snapshot rich scenarios \cite{HanWidebandSPL}, the performance can suffer severely for low SNR, harmonic sources with gaps in spectral energy, and snapshot limited scenarios. Any outliers from a single frequency bin can lead to inaccurate enumeration through the averaging process \cite{CSSM}. 


\section{Proposed Wideband Source Enumeration Scheme for Sparse Arrays}
\label{SecAlgorithm}

\begin{figure}
\begin{center}
\includegraphics[width=0.45\textwidth]{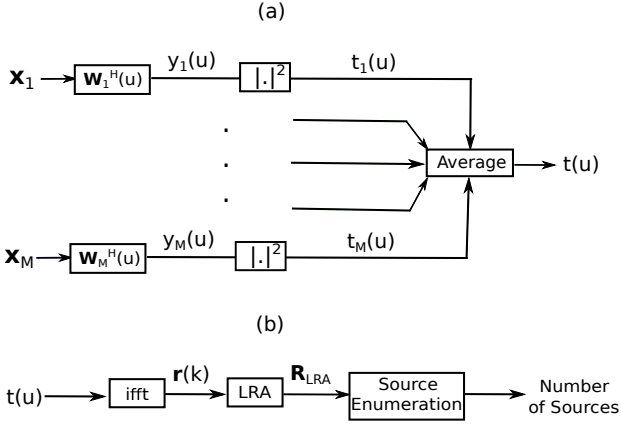} 
\caption{Block diagram for the proposed wideband sparse array source enumeration algorithm. The narrowband spatial periodograms $t_m(u)$ corresponding to each temporal frequency bin $f_1,...,f_M$ are averaged to obtain the wideband periodogram (a) before computing the inverse Fourier transform for the lag redundancy averaged ACM and source enumeration using the proposed information criteria framework (b).}
\label{block}
\end{center}

\end{figure}

The proposed source enumeration scheme contains three major components: wideband spatial periodogram estimation (Fig. \ref{block}a), lag redundancy averaged ACM construction, and wideband source enumeration (Fig. \ref{block}b). 

\subsection{Spatial Periodogram Averaging}
The spatial periodogram averaging (AP) for sparse arrays extends Hinich's wideband beamformer for ULAs, which exploits the frequency diversity of the scanned responses across the signal bandwidth while processing a single ULA \cite{Hinich}. For each frequency bin $f_1,...,f_M$, the array frequency snapshot data in (\ref{snapshots}) are conventionally beamformed via FFT and averaged over all snapshots to estimate the narrowband spatial periodogram $t_{m}(u)$ for frequency $f_m$. The estimated narrowband spatial periodogram $t_m(u)$ is the Fourier transform of the spatial auto-correlation function in (\ref{corr}) that is routinely used to construct the ACM for DOA estimation \cite{PPLiuSPL,LRAperformance}. In wideband processing, only the true source peaks remain fixed in directional cosine $u$ across different frequency bins, while all other sidelobes change their locations in angle as the temporal frequency varies. Averaging the periodograms across frequencies constructively reinforces the energy at the true source locations while all other sidelobes are relatively attenuated
\begin{equation}
t(u) = \frac{1}{ML}\sum_{m=1}^M \sum_{l=1}^L |\textbf{w}_m^H(u) \textbf{x}_l(f_m)|^2,
\end{equation}
where $\textbf{w}_m(u)$ is the conventional beamforming weight vector for steering direction $u$ and frequency $f_m$. Taking inverse spatial Fourier transform of the wideband periodogram $t(u)$ with respect to directional cosine $u$ and normalizing for the co-array weights $\eta(k)$ estimates the spatial correlation function
\begin{equation}
\tilde{{r}}(k) = \frac{\mathscr{F}^{-1}_{u}(t(u))}{\eta(k)},~k = -(P-1),...,(P-1).
\end{equation} 

\subsection{LRA-based Covariance Matrix Augmentation}
An alternative approach to spatial smoothing for the ACM construction is lag redundancy averaging (LRA) \cite{PillaiACM,PillaiORG,LRAORG}. This technique averages lag estimates from different sensor pairs of equal separation to exploit co-array redundancies and then replaces the individual estimates at that lag by their average \cite{LRAperformance,noteLRA}. As a result, the correlation estimates populating the ACM have reduced variances. The LRA-ACM is constructed as
\begin{equation}
\label{LRA-ACM}
\textbf{R}_{\text{LRA}} = \left[ \begin{array}{cccc}
\tilde{{r}}(0) & \tilde{{r}}(-1) & \cdots & \tilde{{r}}(1-P) \\ 
\tilde{{r}}(1) & \tilde{{r}}(0) & \cdots & \tilde{{r}}(2-P) \\ 
\vdots & \vdots & \ddots & \vdots \\ 
\tilde{{r}}(P-1) & \tilde{{r}}(P-2) & \cdots & \tilde{{r}}(0)
\end{array} \right].
\end{equation}The LRA approach constrains a Hermitian Toeplitz ACM from the correlation estimates, although the ACM is not guaranteed positive semi-definite. For the same sparse array data, the LRA-ACM is more computationally efficient than the SS-ACM. Also, the LRA-ACM exploits the second-order statistics, as opposed to the fourth-order statistics in SS-ACM, and thereby making the performance analysis more tractable. For uncorrelated sources impinging on a ULA, the expectation of the squared Frobenius norm corresponding to the error between the LRA covariance matrix and the true covariance matrx is less than the expected error between the true covariance matrix and the SCM \cite{LRAcoherentsignals}. This implies LRA leads to improved accuracy in estimating the eigenvalues of the true covariance matrix \cite{LinebargerPhD}. For finite snapshots, the SS-ACM in (\ref{SS-ACM}) is related to the LRA-ACM by \cite{PPLiuSPL}
\begin{equation}
\textbf{R}_{ss} = \textbf{R}^2_{\text{LRA}}/P.
\end{equation}This equality indicates that $\textbf{R}_{ss}$ and $\textbf{R}_{\text{LRA}}$ share the same eigen-space and that the eigenvalues of $\textbf{R}_{ss}$ are proportional to the square of the eigenvalues of $\textbf{R}_{\text{LRA}}$. For infinite snapshots, the LRA-ACM approaches 
\begin{equation}
\lim_{L \rightarrow \infty} \textbf{R}_{\text{LRA}} = \textbf{A} \bold \Lambda \textbf{A}^H + \sigma_n^2 \textbf{I}_{P\times P},
\end{equation}
where $\textbf{A}$ is the array manifold matrix for an equivalent full ULA with $P$ elements and $\bold \Lambda = \text{diag}([\sigma_1^2, \sigma_2^2, ..., \sigma_D^2])$ is a diagonal matrix containing all source powers \cite{PPLiuSPL}. The asymptotic expression of the LRA-ACM has the same form as the ensemble covariance matrix for a full ULA with $P$ elements. Thus, the magnitudes of the eigenvalues of the LRA-ACM are more appropriate than those of the SS-ACM to use with source enumeration criteria.

\subsection{Wideband Source Enumeration}
The source enumeration criteria processes the magnitudes of the eigenvalues of the LRA-ACM constructed from the wideband periodogram. The eigenvalue magnitudes are assumed to be in a descending order 
\begin{equation}
|\lambda_1| \geq |\lambda_2| \geq ... \geq |\lambda_k| \geq ... \geq |\lambda_P|.
\end{equation} For source enumeration, Rissanen proposed choosing the model order which yields the minimum code length over a range of candidate orders \cite{MDL,WaxKailathSN}. The MDL criterion is the sum of the log-likelihood of the maximum likelihood estimator of the model parameters and a bias correction term penalizing over-fitting of the model order
\begin{equation}
\text{MDL}(q) = - \log \left( \frac{g_q}{a_q}  \right)^{(P-q)L} + \frac{1}{2}q(2P-q)\log L, 
\end{equation} for $L$ snapshots and possible number of sources $q \in [0,..., P-1]$. The functions $g_q = \prod_{j = q+1}^P |\lambda_j|^{1/(P-q)}$ and $a_q = \frac{1}{P-q}  \sum_{j = q+1}^P |\lambda_j|$ are, respectively, the geometric and arithmetic means of the $P-q$ smallest eigenvalues of the SCM. The estimated number of sources is $\hat{D}_{mdl} = \arg \min_q \text{MDL}(q)$. We modify the MDL criterion and apply it on the LRA-ACM for underdetermined source enumeration. The new information criterion, termed MDLgap, is defined as the first-order backward difference of the MDL criterion normalized over the number of snapshots such that

\begin{small}
\begin{eqnarray}
\label{MDlgapcriteria}
 \text{MDLgap}(q) &=& (1/L)(\text{MDL}(q) - \text{MDL}(q-1))\\ \nonumber
&=& -\log \left( \frac{(a_{q-1})^{P-q+1}}{|\lambda_d| (a_{q})^{P-q} } \right) + \frac{P-q+1/2}{L} \log L, 
\end{eqnarray}\end{small}for the candidate number of sources $q = 1,...,P-1$ and the number of snapshots $L$. The estimated number of sources is $\hat{D}_{gap} = \arg \min_q \text{MDLgap}(q)$. 

\subsection{Consistency Proof of MDLgap for Source Enumeration}
This section proves that the MDLgap criterion in (\ref{MDlgapcriteria}) achieves a consistent estimate of the true number of sources when the ensemble covariance matrix (ECM) has equal strength signal eigenvalues. For the general case when the ECM eigenvalues are not equal, the MDLgap criterion will not over-estimate the number of sources for the large snapshot limit.

Pillai and Haber found that the augmented sample Bartlett estimator $\hat{P}_{B}^a(u)$ from an Hermitian Toeplitz ACM is a sum of weighted dependent $\chi^2$-distributed random variables with mean and variance (Eqs. 31, [A.11] in \cite{PillaiACM}),
\begin{eqnarray}
\label{ACMMoments}
E\{\hat{P}_{B}^a(u)\} &=& \textbf{v}_a^H(u) \textbf{R}_a \textbf{v}_a(u) \triangleq P_B(u) \nonumber \\
\operatorname{Var}\{\hat{P}_{B}^a(u)\} &=& \frac{2}{L}\sum \text{Quantities independent of } L, \nonumber
\end{eqnarray}where $\textbf{v}_a(u)$ is the $P\times 1$ steering vector at direction $u$ and $\textbf{R}_a$ is the $P\times P$ ECM for a $P$-element fully populated ULA. Replacing the steering vector $\textbf{v}_a(u)$  by the eigenvectors of the ACM, the Komolgorov strong law of large numbers (Theorem 2.3.10, \cite{SLLN}) guarantees that the eigenvalues of the ACM converge to the ensemble eigenvalues with probability 1 (almost surely) in large snapshot limit \cite{PPLiuSPL}
\begin{equation}
(\lambda_1,..., \lambda_D, \lambda_{D+1}, ..., \lambda_P) \xrightarrow{a. s.} (\ell_1,...,\ell_D, \ell_{D+1}, ...,\ell_P),
\end{equation}
where the signal eigenvalues $(\ell_1,...,\ell_D) = (\sigma_1^2 + \sigma_n^2, ..., \sigma_D^2+\sigma_n^2)$ and the noise eigenvalues $(\ell_{D+1},...,\ell_P) = (\sigma_n^2,...,\sigma_n^2)$. Since the MDLgap criterion is a real-valued continuous function of the eigenvalues $\lambda_1, ...,\lambda_P$, (\ref{MDlgapcriteria}) converges with probability 1 in large snapshot limit
\begin{equation}
\label{MDLgapConverge}
\text{MDLgap}(q) \xrightarrow{a.s}  h(q) = -\log \left( \frac{(a_{q-1})^{P-q+1}}{\ell_q (a_{q})^{P-q} } \right) , 
\end{equation} where the samples eigenvalues $\lambda_1,...,\lambda_P$ are replaced with the ensemble eigenvalues $\ell_1,...,\ell_P$ in $h(q)$ and $a_q = \frac{1}{P-q}  \sum_{i = q+1}^P \ell_i$ is the arithmetic mean of the $P-q$ smallest ensemble eigenvalues. The proof is given by two cases. First, we prove that in the asymptotic limit, MDLgap will never overestimate the number of sources, regardless of the distribution of source power.  Second, we show for the more limited case of equal source eigenvalues, MDLgap will not underestimate the number of sources.   The combination of these two cases proves that when the source eigenvalues are all equal in power, MDLgap is asymptotically consistent.  

$\textit{Case 1: Over-estimation.}$ Let  $\Delta$ be a positive integer with in $1 \leq \Delta < P-D.$ In this case, the ensemble eigenvalues $\ell_D = \sigma_D^2+\sigma_n^2$ and $\ell_{D+\Delta} = \sigma_n^2$. We evaluate the difference
\begin{equation}
\label{C4}
h(D) - h(D+\Delta)= -\log \left( \frac{\left(\sigma_n^2 + \frac{\sigma_D^2}{P-D+1} \right)^{P-D+1}}{(\sigma_n^2)^{P-D+1} + \sigma_D^2(\sigma_n^2)^{P-D}} \right).
\end{equation}Using binomial expansion, the numerator term in the logarithm can be shown greater than the denominator term for any $\sigma_D^2$ and $\sigma_n^2$. As a result, $h(D) -h(D+\Delta)$ is negative with probability 1 for $1 \leq \Delta < P-D$. Since the negativity of (\ref{C4}) does not rely on any assumption about the signal powers, $h(q)$ will not over-estimate the signal subspace dimension in the large snapshot limit. 

$\textit{Case 2: Under-estimation.}$ Let $\Delta$ be a negative integer $1-D \leq \Delta \leq -1.$ We limit our analysis to the case where the signal eigenvalues are all equal $\ell_1 = \ell_2 = \cdots = \ell_D = \sigma_s^2 + \sigma_n^2$, where $\sigma_s^2$ is the average signal power. Under this assumption, we evaluate the following quantity
\begin{eqnarray}
\label{MDLgapdiffCase2}
&~& h(D) - h(D+\Delta) \nonumber \\
 &=& \log \left( \frac{\left(\frac{(1-\Delta)\text{SNR}}{P-D-\Delta+1} +1 \right)^{P-D-\Delta+1}}{\left(\frac{\text{SNR}}{P-D+1}+1 \right)^{P-D+1} \left( \frac{-\Delta \text{SNR}}{P-D-\Delta}+1\right)^{P-D-\Delta}} \right)   \nonumber \\
&=& f(1-\Delta) - f(1) - f(-\Delta),
\end{eqnarray}
where the function $f(x) \triangleq (P-D+x)\log \left(\frac{x\text{SNR}}{P-D+x}+1 \right)$ for all $x \geq 0$ and $\text{SNR} = \sigma_s^2/\sigma_n^2$. The function $f(x)$ can be shown to be monotonically increasing and concave for $x \geq 0$ and $P>D$. Therefore, the fundamental theorem of calculus yields the following inequality
\begin{equation}
f(a+1)-f(a) > f(a+b+1)-f(a+b)
\end{equation} for $a\geq 0$ and $b> 0$. Let $a = 0, ~b = -\Delta$ and observe that $f(0) = 0$, yielding $f(1-\Delta) - f(1) - f(-\Delta)  < 0$, thus implying $h(D) -h(D+\Delta)$ is negative with probability 1 for $1-D \leq \Delta \leq -1$. The negativity of (\ref{MDLgapdiffCase2}) indicates $h(q)$ will not under-estimate the signal subspace dimension when the ensemble signal eigenvalues are all equal. $\blacksquare$

The two cases above prove that when the signal eigenvalues are all equal, $\text{MDLgap}(D) - \text{MDLgap}(D+\Delta)$ is negative with probability 1 for any non-zero integer $ 1-D \leq \Delta < P-D$ in the large snapshot limit. This implies that when the signal eigenvalues are equal, the MDLgap criterion in (\ref{MDlgapcriteria}) consistently estimates the number of planewave sources. Additionally, MDLgap never over-estimates the number of sources for any configuration of source powers. Simulations presented in the next section demonstrate that the requirement on the equal signal eigenvalues for $1-D \leq \Delta \leq -1$ can be relaxed so long as the signal powers fall within limited dynamic range around their average. Identifying the necessary conditions on this dynamic range such that the MDLgap will not under-estimate the number of sources remains an open problem.

\section{Numerical Simulations}
\label{SecSimulations}






This section uses numerical simulations to demonstrate the advantage of the proposed periodogram averaging (AP) based algorithm over the incoherent subspace (ISS) algorithm \cite{WaxKailathSN,HanWidebandSPL,RajAIC} through MDLgap for wideband source enumeration. For all simulations in this section, the sources are assumed uncorrelated and temporally white and complex Gaussian occupying the same bandwidth of 40 Hz around the central frequency of 100 Hz. The wideband sources are decomposed evenly into 41 narrowband components via FFT with equal amplitudes for the temporal spectrum within the bandwidth. As a benchmark, we compare all simulations against the narrowband (NB) case reflecting the time-bandwidth product of the wideband sources. This means the narrowband sources has 41 times more snapshots than the wideband sources. 

Note that the proposed wideband source enumeration scheme applies in general to any sparse array geometry with a contiguous region in its difference co-array, such as coprime arrays and nested arrays \cite{PPCSA}\cite{PPNested}. For demonstration purposes, we use a MRA with 6 sensors at locations $[1,2,5,6,12,14]d$. This array offers a contiguous coarray region spanning $k \in [-13,13]$. The fundamental inter-element spacing of the MRA assumes to be $d = \lambda/2$, where $\lambda$ is the spatial wavelength at the central frequency $f =$ 100 Hz. The sensor SNR level is defined as the ratio between the power of each source signal to the noise power at a single sensor. The noise is assumed both temporally and spatially white and complex Gaussian occupying the same bandwidth as the sources, uncorrelated from the sources and also in between sensors. The following simulations consider two scenarios focusing on different perspectives. The first is an over-determined scenario validating the algorithm's capability in enumerating closely spaced sources. The second is an under-determined scenario validating the algorithm's capability in enumerating more sources than sensors. 

As a comparison to the proposed MDLgap criterion, we use a second-order statistics of eigen-values (SORTE) criterion for source enumeration, given its capability in enumerating more wideband sources than sensors using incoherent processing \cite{HanWidebandSPL}. The SORTE criterion is a relatively new cluster detection criterion that could be used for source enumeration \cite{SORTE}. This criterion is based on a gap measure of the eigen-values of the covariance matrix, defined by 
\begin{equation}
\text{SORTE}(q) = \frac{var ( \{\nabla \lambda_i\}_{i=q+1}^{P-1} )}{var ( \{\nabla \lambda_i\}_{i=q}^{P-1} )},
\end{equation}
for the possible number of sources $q = 1,...,P-2$. The criterion is set to infinity when its denominator $var ( \{\nabla \lambda_i\}_{i=q}^{P-1} ) = 0$ for any $q$. The quantity
\begin{equation}
var ( \{\nabla \lambda_i\}_{i=q}^{P-1} ) = \frac{1}{P-q} \sum_{i=q}^{P-1} \left(\nabla \lambda_i -\frac{1}{P-q} \sum_{j=q}^{P-1} \nabla \lambda_j \right)^2, 
\end{equation} averages the variances corresponding to the smallest $P-q$ eigen-value gaps, where the gap $\nabla \lambda_i = \lambda_i -\lambda_{i+1}$. The detected source number is $\hat{D}_{\text{SORTE}} = \arg \min_q \text{SORTE}(q)$.

\subsection{Resolving two closely-spaced sources}

\begin{figure}
 \centerline{\includegraphics[width=9.5cm]{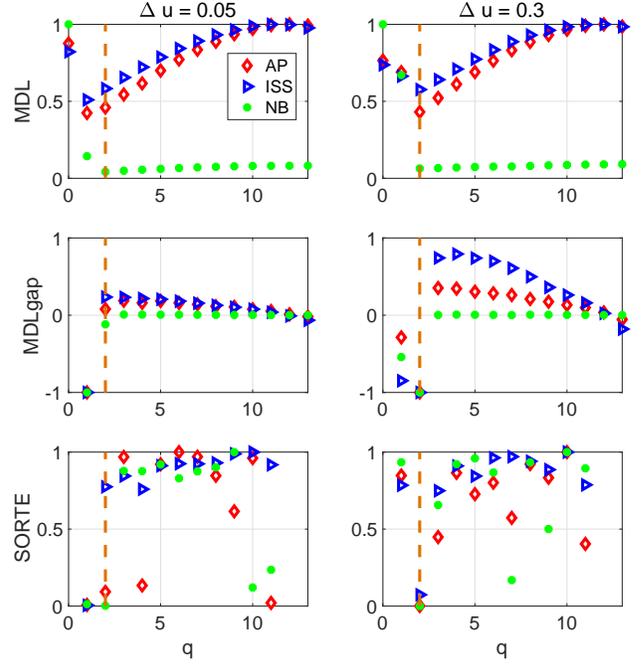}}
\caption{Comparing the sample realizations of MDL, MDL-gap and SORTE criteria for 2 uncorrelated sources with DOAs $u = [0, 0.05]$ on the left column and $u = [0,0.3]$ on the right column. All simulations assume equal power sources with sensor level SNR = 0 dB and 3 snapshots per sensor.}
\label{SampleRealization2sources}
\end{figure}

Fig. \ref{SampleRealization2sources} compares
the sample realizations of the MDL, MDLgap and SORTE criteria as a function of possible number
of sources. All information criteria are normalized by their maximum magnitudes
respectively for demonstration purpose. For simplicity, all sources
are assumed equal power with sensor level SNR = 0 dB. All wideband simulations assume 3 snapshots per sensor and equivalently, 123 snapshots/sensor for the narrowband sources. The left panels simulate two closely arrived sources with one source from broadside and the other from $u = 0.05$. The right panels simulate two sources further separated with one source from broadside and the other from $u = 0.3$. All panels use orange vertical dashed lines to indicate the true number of sources $q = 2$, where all information criteria should show minima for correct source enumeration. The left panels imply that when two sources are close in space, all information criteria struggle to enumerate them with minima occurred at $q = 1$. When the sources are further apart, the right panels imply all information criteria are able to enumerate them correctly with minima occurred at $q=2$.

\begin{figure}
  \centerline{\includegraphics[width=8.5cm]{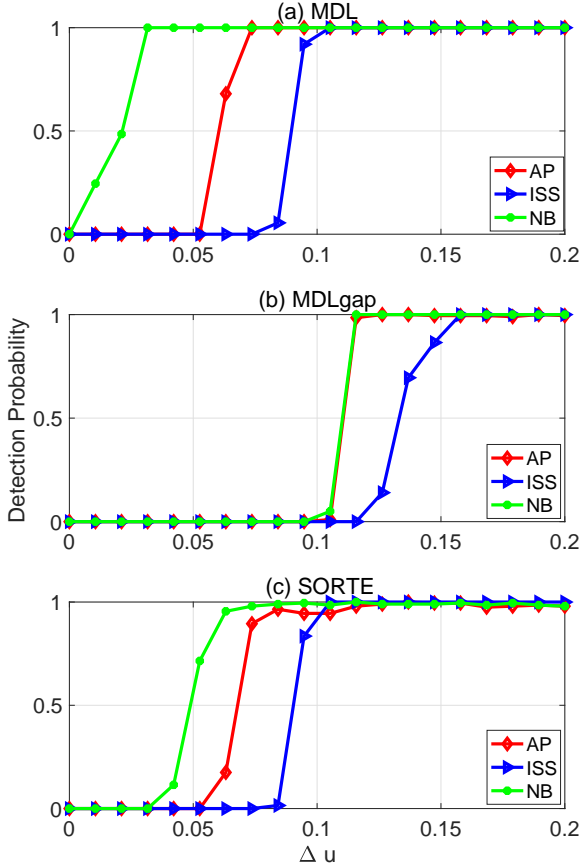}}
\caption{Comparing the probability of detection using MDL, MDLgap and SORTE as a function of the spacing between 2 uncorrelated equal power sources with SNR = 0 dB. The simulations for wideband sources use 3 snapshots/sensor and the equivalent narrowband sources use 123 snapshots per sensor.}
\label{DetectAgaintSeparation}
\end{figure}

Fig. \ref{ProbDetectionAgainstSnapshots} evaluates the probability of correctly estimating $D=2$ as a function of source separation using the MDL, MDLgap and SORTE criteria. The detection probability is calculated as the number of Monte Carlo trials correctly estimating $\hat{D}= 2$, normalized over a total of 200 trials. Overall, all algorithms have a better chance of correct enumeration as the two sources move away from each other. MDL and SORTE have comparable performance in enumerating closely spaced sources, which is slightly better than MDLgap. The proposed AP algorithm is capable of resolving more closely spaced sources than the ISS algorithm for each of the three information criteria.

\subsection{Enumerating more sources than sensors}


This section validates the advantages the proposed AP algorithm offers over the ISS algorithm in enumerating more sources than sensors.  We again use the same 6-element MRA as in the previous section, but with 9 sources: 1 at broadside, 4 uniformly spaced in $\theta = (90^o, 135^o]$ and the other 4 uniformly spaced in $u = (0,0.7]$. All sources are uncorrelated and equal-power with sensor level SNR of 0 dB. 

\begin{figure}
  \centerline{\includegraphics[width=9.5cm]{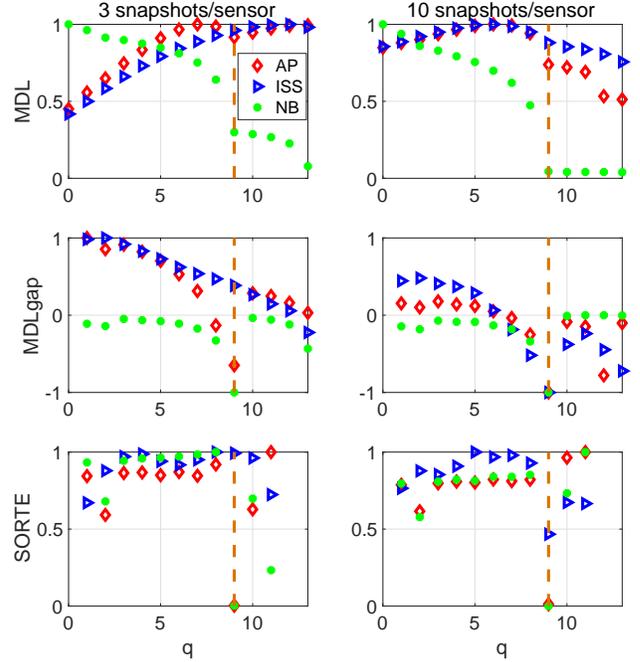}}
\caption{Comparing the sample realizations of MDL, MDL-gap criteria and SORTE criteria for 9 uncorrelated sources. All simulations assume equal power sources with sensor level SNR = 0 dB and 3 snapshots per sensor for the left panels and 10 snapshots per sensor for the right panels.}
\label{SampleRealization9sources}
\end{figure}

Fig.~\ref{SampleRealization9sources} compares the sample realizations of the criteria as a function of possible number of sources, where all information criteria are normalized by their maximum magnitudes respectively. The simulations on the left panels use 3 snapshots/sensor for the wideband source and equivalently, 123 snapshots/sensor for the narrowband source. The simulations on right panels use 10 snapshots/sensor for the wideband source and equivalently, 410 snapshots/sensor for the narrowband source. For all panels, the true number of sources D = 9 is indicated by orange dashed lines. The top left panel shows that when the number of sources D = 9 exceeds the number of sensors N = 6, none of the algorithms exhibit a minimal MDL value at $\hat{D}$ = 9. The middle and bottom left panels show that the AP and NB algorithms show minimal MDLgap and SORTE values at $\hat{D}$ = 9. However, the ISS algorithm is not able to estimate $\hat{D}$ = 9 using either criteria at this modest snapshots level. When the number of snapshots increases, the top right panel shows the MDL still fails to estimate $\hat{D}$ = 9. However, the middle and bottom right panels show that all approaches using the MDLgap and SORTE criteria are able to correctly estimate the source number $\hat{D}$ = 9. These simulations imply that the AP algorithm is capable of enumerating more sources than sensors in relatively few snapshots using MDLgap and SORTE. However, the ISS approach might require relatively large number of snapshots to achieve an accurate enumeration of more sources than sensors. 

\begin{figure}
\includegraphics[width=9cm]{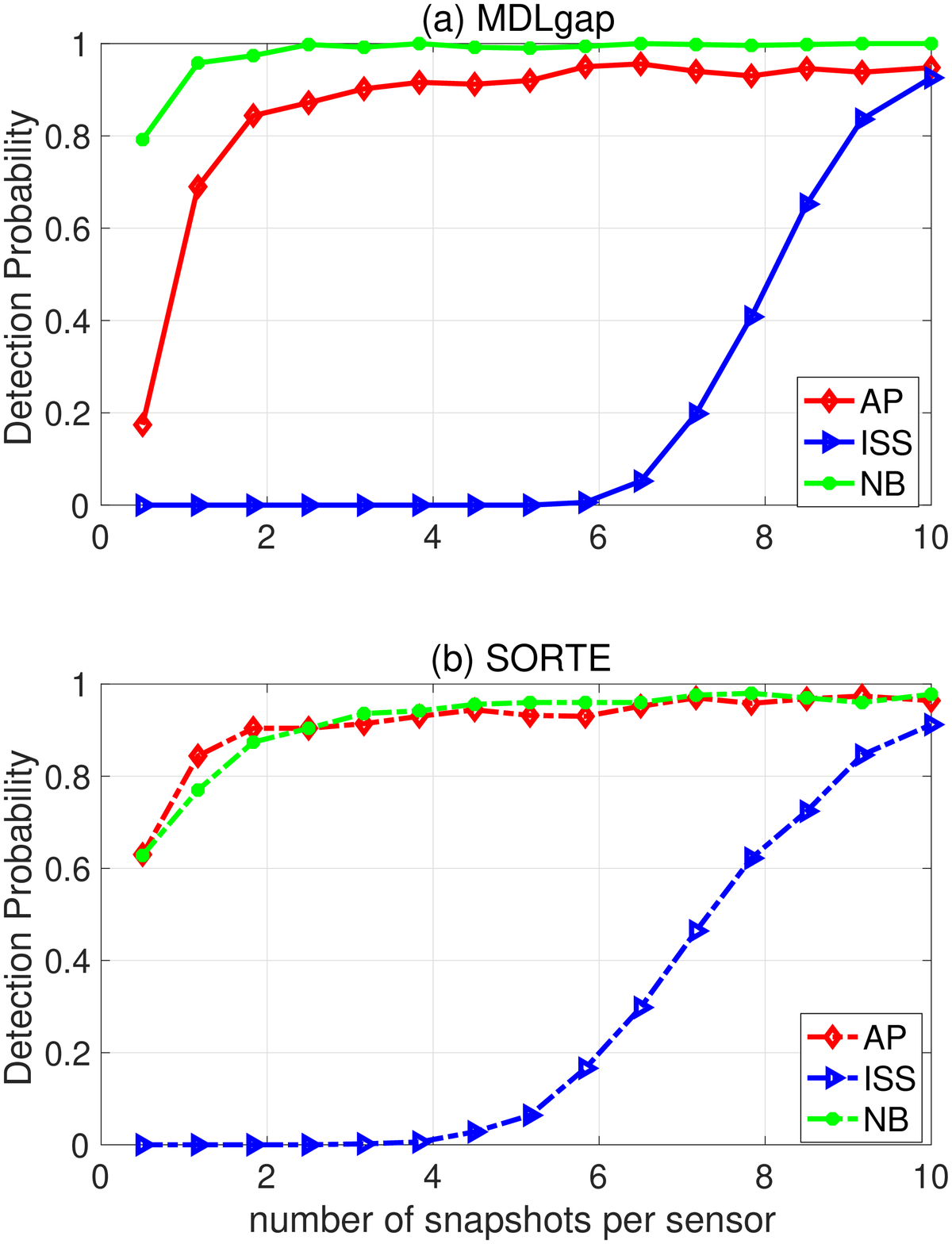} 
\caption{Comparing the probability of correctly enumerating the number of sources using the MDL-gap criterion  for different approaches as a function of the number of snapshots per sensor for fixed sensor level SNR = 0 dB (a) and as a function of sensor level SNR for a fixed 5 snapshots per sensor (b). There are 9 equal power sources impinging on the 6-element MRA.}
\label{ProbDetectionAgainstSnapshots}
\end{figure}

To rigorously validate the performance of the AP algorithm in enumerating more sources than sensors, Fig. \ref{ProbDetectionAgainstSnapshots} evaluates the probability of correctly estimating the number of sources as a function of snapshots/sensor using the MDLgap and SORTE criteria. The detection probability is calculated as the number of Monte Carlo trials correctly estimating $\hat{D}= 9$ sources, normalized over a total of 500 trials. The sensor level SNRs are the same of 0 dB for all 9 sources. Panel (a) shows that using the MDLgap criteria, AP requires many fewer snapshots to achieve the same detection probability as the ISS approach. In contrast, ISS requires 6 snapshot/sensor to start detecting all sources and 10 snapshots/sensor to achieve a detection probability above $90\%$. Panel (b) shows that using the SORTE criteria, AP achieves detection probability above $80\%$ for 1 snapshot/sensor and converges fast to above $95\%$ as the number of snapshots increases. In contrast, ISS requires 4 snapshots/sensor to start detecting all sources and 10 snapshots/sensor to achieve a detection probability above $90\%$. The AP is only slightly worse than the NB case, suggesting that the AP pays only a slight penalty to combine uncorrelated measurements across frequency bands relative to the NB algorithm given an equal number of measurements in a single frequency band.  

\begin{figure}
\begin{center}
\includegraphics[width=9cm]{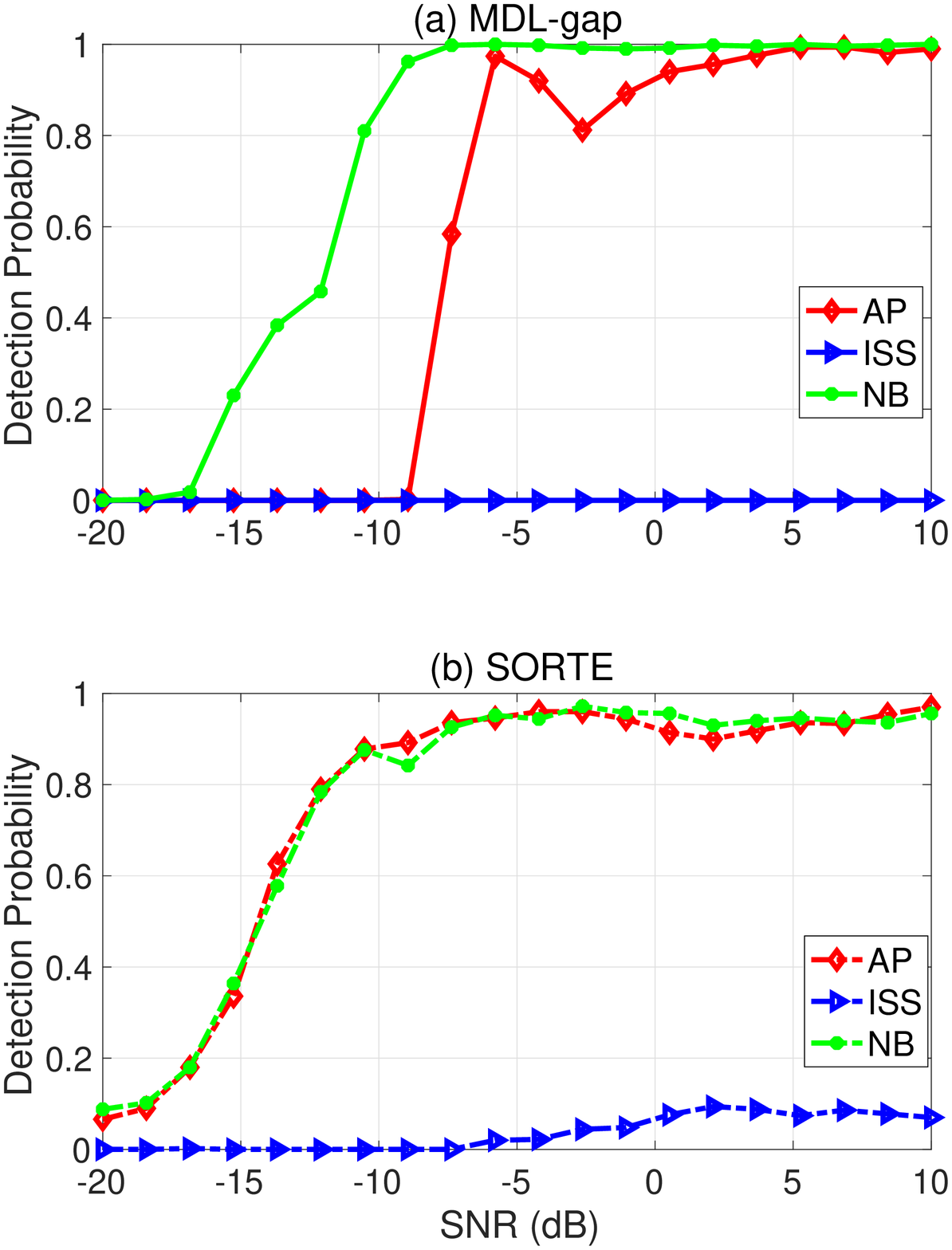} 
\caption{Comparing the probability of correctly enumerating the number of sources using the MDL-gap criterion  for different approaches as a function of the number of snapshots per sensor for fixed sensor level SNR = 0 dB (a) and as a function of sensor level SNR for a fixed 5 snapshots per sensor (b). There are 9 equal power sources impinging on the 6-element MRA.}
\label{ProbDetectionAgainstSNR}
\end{center}
\end{figure}

Fig. \ref{ProbDetectionAgainstSNR} evaluates the detection probability as a function of sensor level SNR using the MDLgap and SORTE criteria. The number of snapshots/sensor is fixed as 5 for the wideband approaches and 205 for the equivalent NB case. Panel (a) shows that using the MDLgap criteria, the NB approach requires the lowest SNR level to start correctly detecting all sources. AP requires SNR level of -9 dB to start enumerating all sources. ISS is not able to enumerate all sources for all SNRs due to the low number of snapshots available. Panel (b) shows that using the SORTE criteria, AP has almost identical detection probability as the narrowband case, achieving $90\%$ for SNR above -10 dB. Again, ISS is barely able to enumerate all sources due to the low number of snapshots available.

The simulations above indicate that the proposed periodogram averaging (AP) algorithm offers clear advantages over the ISS algorithm in enumerating more wideband sources than sensors. These advantages is more prominent when the sources are closely spaced, in low SNR and relatively few snapshots scenarios. Through the proposed MDLgap criteria as opposed to the traditional MDL criteria, it's possible to enumerate more sources than sensors using current popular sparse array geometries such as MRA, coprime and nested arrays. 

\section{Conclusion}
\label{conclusion}
This paper proposes the MDLgap information criterion for wideband source enumeration through spatial periodogram averaging. The proposed algorithm applies to any sparse array geometries for underdetermined scenarios with more sources than sensors. Simulations demonstrate that reinforcing the sources' spectral information by averaging the narrowband periodograms across frequency greatly improves the source enumeration performance when compared with approaches incoherently averaging the information criteria across frequency, especially in low snapshot scenarios. Our future research efforts hope to strengthen the asymptotic consistency proof of the MDLgap criteria by relaxing the requirement that the ensemble covariance matrix has equal signal eigenvalues.



%

\ifCLASSOPTIONcaptionsoff
  \newpage
\fi




\begin{thebibliography}{1}
\bibitem{Johnson}
D.H. Johnson and D.E. Dudgeon, \textit{Array Signal Processing: Concepts and Techniques.} Simon \& Schuster, New York, NY, 1992.

\bibitem{MRA}
A. Moffet, ``Minimum-redundancy linear arrays,'' \textit{IEEE Trans. Antennas Propag.}, vol. AP-16, no. 2, pp. 172-175, 1968.

\bibitem{PPCSA} 
P.P. Vaidyanathan and P. Pal, ``Sparse sensing with co-prime samplers and arrays,'' \textit{IEEE Trans. Signal Process.}, vol. 59, no. 2, pp. 573-586, 2011.

\bibitem{PPNested}
P. Pal and P.P. Vaidyanathan, ``Nested arrays: a novel approach to array processing with enhanced degrees of freedom,'' \textit{IEEE Trans. Signal Process.}, vol. 58, no. 8, pp. 4167-4181, 2010. 

\bibitem{CSADetectionICASSP} 
K. Adhikari and J.R. Buck, ``Gaussian signal detection by coprime sensor arrays,'' in \textit{Proc. IEEE Int. Conf. Acoust., Speech, Signal Process.}, pp. 2379-2383, 2015.

\bibitem{kaushallyaECSA} 
K. Adhikari, J.R. Buck and K.E. Wage, ``Extending coprime sensor arrays to achieve the peak side lobe height of a full uniform linear array,'' \textit{EURASIP J. Adv. Signal Process.}, vol. 1, pp. 1-17, 2014.

\bibitem{YangSAM2016}
Y. Liu and J.R. Buck, ``Spatial spectral estimation using a coprime sensor array with the min processor,'' in \textit{Proc. 9th IEEE Sensor Array and Multichannel Signal Processing Workshop (SAM)}, pp. 1-5, 2016.

\bibitem{PillaiACM}
S.U. Pillai and F. Haber, ``Statistical analysis of a high resolution spatial spectrum estimator utilizing an augmented covariance matrix,'' \textit{IEEE Trans. Acoust., Speech Signal Process.,}, vol. 35, no. 11, pp. 1517-1523, 1987.

\bibitem{PPCSAMUSIC}
P. Pal and P.P. Vaidyanathan, ``Coprime sampling and the MUSIC algorithm,'' in \textit{Proc. IEEE Digital Signal Processing Workshop and IEEE Signal Processing Education Workshop (DSP/SPE)}, pp. 289-294, 2011.

\bibitem{Aminmultiple}
E. BouDaher, Y. Jia, F. Ahmad and M.G. Amin, ``Multi-frequency co-prime arrays for high-resolution direction-of-arrival estimation,'' \textit{IEEE Trans. Signal Process.}, vol. 63, no. 14, pp. 3797-3808, 2015.

\bibitem{AIC}
H. Akaike, ``A new look at the statistical model identification,'' \textit{IEEE Trans. Autom. Control}, vol. 19. no. 6, pp. 716-723, 1974.

\bibitem{MDL}
J. Rissanen, ``Modeling by shortest data description,'' \textit{Automatica}, vol. 14, pp. 465-471, 1978.

\bibitem{WaxKailathSN}
M. Wax and T. Kailath, ``Detection of signals by information theoretic criteria,'' \textit{IEEE Trans. Acoust. Speech Signal Process.}, vol. 33, no. 2, pp. 387-392, 1985.

\bibitem{PPLiuSPL}
C.-L. Liu and P.P. Vaidyanathan, ``Remarks on the spatial smoothing step in Co-array MUSIC,'' \textit{IEEE Signal Process. Lett.}, vol. 22, no. 9, pp. 1438-1442, 2015.


\bibitem{BragCox}
A.B. Baggeroer and H. Cox, ``Passive sonar limits upon nulling multiple moving ships with large aperture arrays,'' in \textit{Proc. IEEE Asil. Conf. on Signals, Syst., Comput}, pp. 103-108, 1999.

\bibitem{Cox}
H. Cox, ``Adaptive beamforming in non-stationary environments,'' in \textit{Proc. IEEE Asil. Conf. on Signals, Syst., Comput}, pp. 431-438, 2002.

\bibitem{RajAIC}
R.R. Nadakuditi and A. Edelman, ``Sample eigenvalue based detection of high-dimensional signals in white noise using relatively few samples,'' \textit{IEEE Trans. Signal Process.}, vol. 56, no. 7, pp. 2625-2638, 2008.

\bibitem{kritchman2008}
S. Kritchman and B. Nadler, ``Determining the number of components in a factor model from limited noisy data,'' \textit{Chemom. Intell. Lab. Syst.}, vol. 94, no. 1, pp. 19-32, 2008. 


\bibitem{Wishart}
J. Wishart, ``The generalized product moment distribution in samples from a normal multivariate population,'' \textit{Biometrika}, vol. 20, A, pp. 32-52, 1928.

\bibitem{HanWidebandSPL}
K. Han and A. Nehorai, ``Wideband Gaussian source processing using a linear nested array,'' \textit{IEEE Signal Process. Lett.}, vol. 20, no. 11, pp. 1110-1113, 2013.

\bibitem{Hinich}
M.J. Hinich, ``Processing spatially aliased arrays,'' \textit{J. Acoust. Soc. Am.}, vol. 64, no. 3, pp. 792-794, 1978. 

\bibitem{CSSM}
H. Wang and M. Kaveh, ``Coherent signal-subspace processing for the detection and estimation of angles of arrival of multiple wide-band sources,'' \textit{IEEE Trans. Acoust. Speech Signal Process.,} vol. 33, no. 4, pp. 823-831, 1985. 


\bibitem{LRAperformance}
M.A. Doron and A.J. Weiss, ``Performance analysis of direction finding using lag redundancy averaging,'' \textit{IEEE Trans. Signal Process.}, vol. 41, no. 3, pp. 1386-1391, 1993.


\bibitem{PillaiORG}
S.U. Pillai, Y. Bar-Ness and F. Haber, ``A new approach to array geometry for improved spatial spectrum estimation,'' \textit{Proc. IEEE}, vol. 73, no. 10, pp. 1522-1524, 1985.

\bibitem{LRAORG}
S.Y. Kung, C.K. Lo and R. Foka, ``A Toeplitz approximation approach to coherent source direction finding,'' in \textit{Proc. IEEE Int. Conf. Acoust., Speech, Signal Process.}, pp. 193-196, 1986.

\bibitem{noteLRA}
K.C. Indukumar and V.U. Reddy, ``A note on redundancy averaging,'' \textit{IEEE Trans. Signal Process.}, vol. 40, no. 2, pp. 466-469, 1992.



\bibitem{LRAcoherentsignals}
D.A. Linebarger and D.H. Johnson, ``The effect of spatial averaging on spatial correlation matrices in the presence of coherent signals,'' \textit{IEEE Trans. Acoust. Speech Signal Process.}, vol. 38, no. 5, pp. 880-884, 1990.

\bibitem{LinebargerPhD}
D. Linebarger, ``Parametric and non-parametric methods of improving bearing estimates in narrow-band passive sonar systems,'' \textit{Ph.D. dissertation}, Rice University, Houston, TX, July 1986.

\bibitem{SLLN}
P.K. Sen and J.M. Singer, \textit{Large Sample Methods in Statistics.} Chapman \& Hall Inc, London, UK, 1993.

\bibitem{SORTE}
Z. He, A. Cichocki, S. Xie and K. Choi, ``Detecting the number of clusters in n-way probabilistic clustering,'' \textit{IEEE Trans. Pattern Anal. Mach. Intell.,} vol. 32, no. 11, pp. 2006-2011, 2010. 

%
%
%
%
%


\end{thebibliography}
%

\bibliographystyle{IEEEtran}

\end{document}